# Strong correlations generically protect $d$-wave superconductivity against disorder

Shao Tang,[1] V. Dobrosavljević,[1] and E. Miranda[2]

[1]*Department of Physics and National High Magnetic Field Laboratory,*
*Florida State University, Tallahassee, Florida 32306, USA*
[2]*Instituto de Física Gleb Wataghin, Campinas State University,*
*Rua Sérgio Buarque de Holanda, 777, CEP 13083-859, Campinas, Brazil*

We address the question of why strongly correlated $d$-wave superconductors, such as the cuprates, prove to be surprisingly robust against the introduction of non-magnetic impurities. We show that, very generally, both the pair-breaking and the normal state transport scattering rates are significantly suppressed by strong correlations effects arising in the proximity to a Mott insulating state. We also show that the correlation-renormalized scattering amplitude is generically enhanced in the forward direction, an effect which was previously often ascribed to the specific scattering by charged impurities outside the copper-oxide planes.

*Introduction.*— For many classes of unconventional superconductors, such as the cuprates [1–4], heavy fermion superconductors [5], organic materials [6, 7] and iron pnictides [8], electronic interactions are believed to be essential. Among the many puzzling features of these systems is their behavior in the presence of disorder. In weakly interacting $d$-wave superconductors, Abrikosov-Gor'kov (AG) theory predicts that a tiny amount of non-magnetic impurities should bring the transition temperature $T_c$ to zero. In the case of the cuprates, however, experiments have shown that these $d$-wave superconductors are very robust against disorder [3, 9–11]. This feature was frequently ascribed to scattering by charged off-plane impurities, which is mostly in the forward direction (see, e.g., [12]). The puzzle was partially clarified, however, once strong electronic interactions were shown to give rise to the impurity screening effects seen in these experiments, as captured by the Gutzwiller-projected wave function [13–17]. Despite this progress, it would be desirable to understand both qualitatively and quantitatively whether disorder screening has any significant influence on $T_c$ as well as on the normal state transport properties.

The transition temperature in the under-doped region of the hole-doped cuprates is believed to be influenced by phase fluctuations, various types of competing orders (such as charge- and spin-density waves), stripe formation, etc. Consequently, impurities act as nucleations centers, which complicates the analysis considerably. In the over-doped region, however, $T_c$ is dominated by the superconducting gap opening, thus offering a particularly favorable window into the interplay between disorder and interactions.

In the presence of impurities, the strongly correlated state readjusts itself and creates a renormalized disorder potential. In the dilute limit, the AG theory can be extended to describe the effect of this renormalized potential on $T_c$ degradation and transport properties. We will describe in this Letter how electronic interactions lead to a much slower decrease of $T_c$ as compared to the weak-coupling theory. Our results demonstrate that (i) this effect is intrinsically tied to the proximity to the Mott insulating state, although it is significant even above optimal doping; (ii) the doping dependence of normal state resistivity is different from that of the pair-breaking scattering rate, which governs $T_c$; and (iii) the softening of the disorder potential by interactions leads to a strong enhancement of the forward scattering amplitude.

*Model and method.*—We start with the $t-J$ model on a cubic lattice in $d$ dimensions with dilute nonmagnetic impurities

$$H = -t\sum_{\langle ij\rangle\sigma} c_{i\sigma}^\dagger c_{j\sigma} + J\sum_{\langle ij\rangle} \mathbf{S}_i\cdot\mathbf{S}_j + \sum_i (\varepsilon_i - \mu_0)n_i, \quad (1)$$

where $c_{i\sigma}^\dagger$ ($c_{i\sigma}$) is the creation (annihilation) operator of an electron with spin projection $\sigma$ on site $i$, $t$ is the hopping matrix element between nearest neighbors, $J$ is the superexchange coupling constant between nearest-neighbor sites, $n_i = \sum_\sigma c_{i\sigma}^\dagger c_{i\sigma}$ is the number operator, $\mu_0$ is the chemical potential. The no-double-occupancy constraint ($n_i \leq 1$) is implied. We work in units such that $\hbar = k_B = a = 1$, where $a$ is the lattice spacing and the total number of lattice sites is $V$. For definiteness, we will set $J = t/3$. The impurities are taken into account through a random on-site potential described by $\varepsilon_i$. We use a model of disorder in which we set the potential $\varepsilon_i = t$ and randomly place the impurities on lattice sites with $n$ impurities per unit volume and no correlations between their positions. Note that this model assumes random non-magnetic scattering but does not describe the removal of magnetic ions. We will focus on the two-dimensional case relevant to the cuprates, but our results are easily generalizable to higher dimensions with few modifications.

We proceed with $U(1)$ slave boson theory, details of which can be found in [4, 18–21]. Briefly, it starts with the replacement $c_{i\sigma}^\dagger \to f_{i\sigma}^\dagger b_i$, where $f_{i\sigma}^\dagger$ and $b_i$ are auxiliary fermionic (spinon) and bosonic (slave boson) fields. This substitution is faithful if the constraint $n_i \leq 1$ is replaced by $\sum_\sigma f_{i\sigma}^\dagger f_{i\sigma} + b_i^\dagger b_i = 1$. The latter is enforced through Lagrange multiplier fields $\lambda_i$ on each site. The $J$ term is then decoupled through additional Hubbard-Stratonovitch bosonic fields in the particle-particle ($\Delta_{ij}$) and particle-hole ($\chi_{ij}$) channels. The auxiliary bosonic fields are all treated in the saddle-point approximation, which here is *spatially inhomogeneous* due to the presence of disorder: $\langle b_i\rangle = r_i$, which governs the local quasi-particle residue $Z_i = r_i^2$, $\langle \lambda_i \rangle$ (we will denote it simply by $\lambda_i$) which renormalizes the site energies and $\chi_{ij} = \sum_\sigma \left\langle f_{i\sigma}^\dagger f_{j\sigma} \right\rangle$ and $\Delta_{ij} = \left\langle f_{i\uparrow} f_{j\downarrow} - f_{i\downarrow} f_{j\uparrow} \right\rangle$, which describe, respectively, the strength of a spinon singlet and the pairing amplitude across the corresponding bonds. We also made the change $J \to \widetilde{J} = \frac{3}{8}J$. This choice is made so that the saddle-point approximation of the above multi-channel Hubbard-Stratonovitch transformation coincides with the mean-field results [4]. We note, however, that the usual choice $\widetilde{J} = \frac{1}{4}J$ would give rise to hardly noticeable changes in the numerical results. We stress that the $f$-electrons mentioned throughout the text are only auxiliary fermions, usually called spinons, rather than the physical electrons. They are related at the saddle point by $c_{i\sigma}^\dagger = r_i f_{i\sigma}^\dagger$. Note that the non-trivial effects of this work come from the self-consistent spatial readjustments of the condensed fields to the disorder potential.

In the clean limit ($\varepsilon_i = 0$) and in the saddle-point approximation, the bosonic fields are spatially uniform: $r_i = r_0$, $\lambda_i = \lambda_0$, $\chi_{ij} = \chi \Gamma_s(i,j)$ and $\Delta_{ij} = \Delta_0 \Gamma_d(i,j)$. Here, $\Gamma_{s,d}(i,j)$ are the real space cubic harmonics which, in **k**-space, are given by $\Gamma_s(\mathbf{k}) = 2(\cos k_x + \cos k_y)$ and $\Gamma_d(\mathbf{k}) = 2(\cos k_x - \cos k_y)$. As the doping level (measured with respect to half-filling) $x = 1 - \sum_i n_i/V = r_0^2$ is increased, the slave boson condensation temperature $T_b$ increases monotonically from zero whereas the $\Delta$ field condenses at a transition temperature $T_\Delta$ which decreases monotonically from a finite value at $x = 0$ to zero at an upper doping level $x_{max}$ [4, 20]. The two curves meet at optimal doping $x_{opt}$. The dome below the two curves is the superconducting dome. Our focus in this paper is on the overdoped region $x > x_{opt}$, in which the superconducting transition temperature $T_c = T_\Delta < T_b$.

Within this spatially inhomogeneous theory, we are able to perform a complete quantitative calculation of the *effective* disorder potential. Details have been explained elsewhere [17]. Here, we will focus on the effects of disorder on the superconducting transition temperature $T_c$ and on transport properties in the correlated normal state for $T \gtrsim T_c$ in the overdoped region. For this purpose, we can set $\Delta_{ij} = 0$. Moreover, in this range of temperature and dopings the other bosons, $r_i$, $\lambda_i$ and $\chi_{ij}$, are thoroughly condensed and therefore fairly insensitive to finite temperature effects. We are thus justified in approximating them by their zero-temperature values.

We will focus on the case of weak scattering by dilute impurities $n \ll 1$, where a linear response theory is sufficient. In other words, we calculate the spatial fluctuations of the various condensed fields to first order in the perturbing potential $\varepsilon_i$ [17]. Extensive numerical calculations carried out both in the normal and in the superconducting states have shown that the crucial spatial fluctuations come from the $\lambda_i$ and $r_i$ fields whereas fluctuations of $\chi_{ij}$ play only a negligible role [17]. We will thus simply fix $\chi_{ij}$ at its clean limit value $\chi$ while allowing for the full self-consistent spatial adjustment of the $\lambda_i$ and $r_i$ fields to the disordered situation.

Given this setup, the superconducting transition at $T_c$ corresponds to the formation of the order parameter $\Delta_{ij} = \langle f_{i\uparrow} f_{j\downarrow} - f_{i\downarrow} f_{j\uparrow} \rangle$. The condensing $f$-electrons, on the other hand, are governed, in the clean limit by a dispersion relation renormalized by the slave boson fields $r_0$ and $\chi$, $\widetilde{h}(\mathbf{k}) \equiv -(tx + \widetilde{J}\chi)\Gamma_s(\mathbf{k})$ and a renormalized chemical potential $\mu_0 - \lambda_0 \equiv -\nu_0$. *This theory, therefore, describes a BCS-type condensation of the f-electrons*. In the presence of disorder, the various fields will readjust themselves. The effect of dilute identical non-magnetic impurities on $T_c$ will therefore be captured within the Abrikosov-Gor'kov (AG) theory [22]. In that theory, the only input needed is the scattering $T$-matrix due to a *single* impurity. For that purpose, we place a single impurity at the lattice origin $\varepsilon_i = t\delta_{i,0}$. Crucially, however, the $\lambda_i$ and $r_i$ fields will differ from their clean-limit value *inside an extended region around the impurity*, not only at the origin. The effective $T$-matrix will thus reflect this non-trivial rearrangement. As shown in reference [17], the impurity potential is "healed" within a length scale of a few lattice parameters, the so-called healing length. Furthermore, it was shown that the healing process/length is strongly influenced by electronic correlations and 'Mottness', even up to dopings $x \approx 0.3$. Therefore, as will be shown, the effective scattering will be strongly suppressed relative to the non-correlated case.

We also look at the transport properties in the normal state around $T_c$. Again, the AG analysis can be straightforwardly applied in our case. The relevant input for the calculation of the resistivity is the *physical electron* scattering $T$-matrix for a single impurity.

A straightforward calculation up to first order in the impurity potential gives the $T$-matrix in momentum space for $f$ fermions and physical ($e$) electrons, respectively, as [23]

$$\langle \mathbf{k} | T^f | \mathbf{k}' \rangle = xt \left[ \frac{h(\mathbf{k}) + h(\mathbf{k}') - \Pi(\mathbf{k}' - \mathbf{k})}{\lambda_0 - \frac{\lambda_0}{2d}\Gamma_s(\mathbf{k}' - \mathbf{k}) - x\Pi(\mathbf{k}' - \mathbf{k})} \right], \quad (2)$$

$$\langle \mathbf{k} | T^e | \mathbf{k}' \rangle = -t \left\{ \frac{\Pi(\mathbf{k}' - \mathbf{k}) + \frac{2\nu_0}{x} + \frac{\widetilde{J}\chi}{tx}[h(\mathbf{k}) + h(\mathbf{k}')]}{\lambda_0 - \frac{\lambda_0}{2d}\Gamma_s(\mathbf{k}' - \mathbf{k}) - x\Pi(\mathbf{k}' - \mathbf{k})} \right\}, \quad (3)$$

where $h(\mathbf{k}) = -t\Gamma_s(\mathbf{k})$ is the bare energy dispersion,

$$\Pi(\mathbf{k}) \equiv \frac{1 + \Pi^b(\mathbf{k})}{\Pi^a(\mathbf{k})}, \quad (4)$$

with

$$\Pi^a(\mathbf{k}) = \frac{1}{V} \sum_\mathbf{q} \frac{f\left[\widetilde{h}(\mathbf{q}+\mathbf{k})\right] - f\left[\widetilde{h}(\mathbf{q})\right]}{\widetilde{h}(\mathbf{q}+\mathbf{k}) - \widetilde{h}(\mathbf{q})},$$

$$\Pi^b(\mathbf{k}) = \frac{1}{V} \sum_\mathbf{q} \frac{f\left[\widetilde{h}(\mathbf{q}+\mathbf{k})\right] - f\left[\widetilde{h}(\mathbf{q})\right]}{\widetilde{h}(\mathbf{q}+\mathbf{k}) - \widetilde{h}(\mathbf{q})} [h(\mathbf{q}+\mathbf{k}) + h(\mathbf{q})],$$

and $f(x)$ is the Fermi-Dirac function at $T = 0$.

In order to assess the role of electronic correlations we will compare our full results as described above with a corresponding non-correlated system in which $J = 0$. In the latter case, the $T$-matrix is given simply by the lattice Fourier transform of the bare disorder potential, $\langle \mathbf{k} | T_0 | \mathbf{k}' \rangle = \varepsilon(\mathbf{k}' - \mathbf{k}) = t$, and there is no distinction between auxiliary and physical fermions. The two sets of results will be called correlated and non-correlated, respectively. Even at this point, the renormalizations due to strong correlations are clear: the $\mathbf{k}$-dependent factors in Eqs. (2) and (3), which reflect the spatial readjustments of the $r_i$ and $\lambda_i$ fields, make the bare potential "softer" and more non-local. Note also the extra $x$ factor in Eq. (2) as compared to Eq. (3).

At low temperatures, only scattering very close to the Fermi level is relevant. We will thus calculate the $T$-matrices at the Fermi surface. Furthermore, we are interested in the overdoped region, where the Fermi surface anisotropy becomes increasingly less pronounced as the doping increases. Therefore, we will simplify the actual lattice dispersion in favor of an isotropic one corresponding to the continuum limit, $h(\mathbf{k}) \approx -4t + tk^2$. This is equivalent to a bare effective mass $m = 1/2t$ and a renormalized one $m^* \equiv 1/(2tx + 2\tilde{J}\chi)$. Finally, we call $E_F$ and $k_F$ the Fermi energy and momentum for the bare dispersion $h(\mathbf{k})$, respectively, while $\widetilde{E}_F = \frac{m^*}{m} E_F$ is the Fermi energy for the renormalized dispersion of the $f$-fermions.

*Pair breaking parameter.—* Once the scattering matrix has been determined, it is a trivial matter to write down the predictions of the Abrikosov-Gor'kov (AG) theory for the suppression of the superconducting transition temperature $T_c$ [23]

$$\ln \frac{T_{c0}}{T_c} = \psi\left(\frac{1}{2} + \frac{\alpha}{2}\right) - \psi\left(\frac{1}{2}\right), \quad (5)$$

where $T_{c0}$ is the transition temperature in clean limit, $\alpha \equiv 1/(2\pi T_c \tau_{pb})$, and $\tau_{pb}$ is the pair breaking scattering time. The latter is given in the continuum limit by

$$\frac{1}{\tau_{pb}} = \frac{x^2 n m^*}{2\pi} \int_0^{2\pi} d\theta\, g\left[\left|\sin\left(\frac{\theta}{2}\right)\right|\right](1 - \cos 2\theta), \quad (6)$$

where

$$g(y) \equiv \frac{t^2}{\{\rho^* \lambda_0 k_F^2 y^2 g_L(y) + x[1 - 2\rho^* E_F g_L(y)]\}^2}, \quad (7)$$

where $\rho^* = \frac{m^*}{2\pi}$ is the renormalized density of states and

$$g_L(y) \equiv \begin{cases} 1 & y \leq 1, \\ 1 - \sqrt{1 - y^{-2}} & y > 1. \end{cases} \quad (8)$$

The factor of $1 - \cos 2\theta$ comes from the vertex corrections for $d$-wave pairing and can be generalized to other pairing symmetries by changing $\cos 2\theta$ to the corresponding lattice harmonic. The leading behavior for low impurity concentrations is

$$T_c = T_{c0} - \pi/8\tau_{pb}. \quad (9)$$

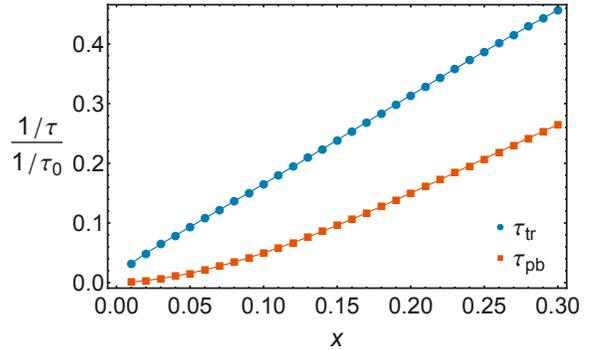

Figure 1: The pair-breaking and transport scattering rates normalized by the non-correlated value $1/\tau_0$ as a function of the doping level.

Fig. 1 shows the ratio of the pair-breaking scattering rate $1/\tau_{pb}$ in the correlated case to the non-correlated one. Note that, for the non-correlated case,

$$\frac{1}{\tau_0} = \frac{nm}{2\pi} \int_0^{2\pi} d\theta\, t^2 (1 - \cos 2\theta) = nmt^2. \quad (10)$$

Clearly, pair breaking is strongly suppressed by electronic correlations. While this suppression is enhanced as the density-driven Mott transition is approached ($x \to 0$), it is still quite significant up to dopings of $x \approx 0.3$. As a result, the $T_c$ degradation is expected to be considerably slower in that case and we expect the $d$-wave superconductivity to be more robust than predicted by the weak coupling theory. Equivalently, the critical impurity concentration $n_c$ at which $T_c$ vanishes is enhanced when compared to the non-correlated case, $5 - 10$ times in the range of dopings from 0.15 to 0.3. We note that this suppression of pair-breaking by the impurities is completely dominated by the $x^2$ dependence of Eq. (6). Indeed, in the whole range of dopings from $\sim 0.01$ to $\sim 0.3$, the product of the effective mass $m^*$ and the angular integral in Eq. (6) varies very little (roughly from 5 to 3). Thus, in a manner very reminiscent of the strong healing of gap fluctuations found in reference [17], here the robustness of $T_c$ can also be attributed to 'Mottness'.

*Transport scattering rate.—* The normal state resistivity is governed by the impurity induced transport scattering rate, which can be evaluated straightforwardly via Eq. (3) to give [23]

$$\frac{1}{\tau_{tr}} = \frac{xnm^*}{2\pi} \int_0^{2\pi} d\theta\, g\left[\left|\sin\left(\frac{\theta}{2}\right)\right|\right](1 - \cos \theta). \quad (11)$$

The non-correlated transport scattering rate defined as

$$\frac{1}{\tau_0^{tr}} = \frac{nm}{2\pi} \int_0^{2\pi} d\theta\, t^2 (1 - \cos \theta), \quad (12)$$

which coincides with the above $1/\tau_0$ for the bare isotropic scattering impurity potential we used. As shown in Fig. 1, the transport rate is also suppressed by electronic correlations and 'Mottness'. In contrast to Eq. (6), however, the dependence is almost linear in $x$. This is because, as before, the product of $m^*$ and the angular integral in Eq. (11) is almost doping independent. As a result, as seen in Fig. 1, for a wide range of doping levels the suppression of the pair-breaking scattering rate is much more significant than the transport one.

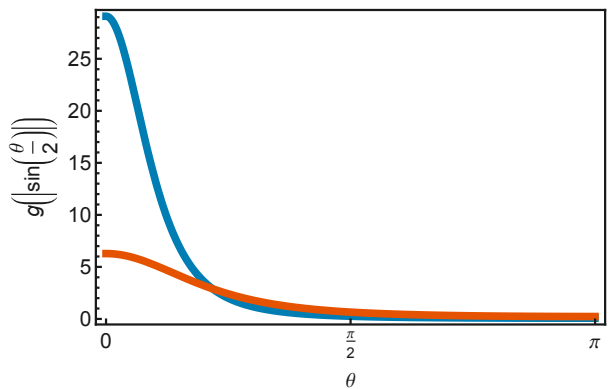

Figure 2: The angular dependence of the renormalized $T$-matrices at $x = 0.15$ (blue) and $x = 0.3$ (red).

*Forward scattering.*— The doping dependence of the scattering rates illustrated in Fig. 1 makes it clear that the dominant effect comes from the explicit $x$ dependence in Eqs. (6) ($\sim x^2$) and (11) ($\sim x$). The $x$-dependence coming from $m^*$ times the angular integrals over the scattering matrices is very weak. However, this does not mean that the angular dependence of the $T$-matrices is not affected by strong correlations, as we will now show.

In Fig. 2 we show, for two doping levels, the angular dependence of the function $g\left[\left|\sin\left(\frac{\theta}{2}\right)\right|\right]$ [defined in Eq. (7)], which is integrated over in Eqs. (6) and (11). This should be compared to the bare impurity result, which is $\sim t^2$ and thus $\theta$-independent. Clearly, there is a large enhancement of forward scattering, indicating a "softening" of the impurity scattering by correlations, even for point-like impurities in the plane.

This function is weighted by $1 - \cos 2\theta$ and $1 - \cos \theta$ in the integrations in Eqs. (6) and (11), respectively. These weight functions amplify the contributions from the regions $\theta \approx \pi/2$ and $\theta \approx \pi$, respectively, which are, however, hardly affected by correlations. As a result, even with the softening of the impurity scattering, the angular integrals are not renormalized significantly in the range from $\sim 0.15$ to $\sim 0.3$, when compared to the non-correlated bare impurity result: $\sim 0.8 - 1$ in Eq. (6) and $\sim 0.3 - 0.5$ in Eq. (11). The conclusion, then, is that strong correlations enhance significantly the forward scattering region even for point-like in-plane impurities, but this is *not* the reason for the robustness of $T_c$ or the resilience of the normal state conductivity.

*Conclusions.*— We have shown how the weak-coupling AG theory of $T_c$ suppression and normal state resistivity by dilute non-magnetic impurities is modified in a strongly correlated metal. Even though the renormalized scattering amplitude is strongly enhanced in the forward direction, the most significant effect comes from the suppression of the electron fluid compressibility by 'Mottness', which is effective even relatively far from the Mott insulating state. Given its simplicity, we suggest that this phenomenon is generic to other systems close to Mott localization.

*Acknowledgments.*— We acknowledge support by CNPq (Brazil) through Grants No. 304311/2010-3 and No. 590093/2011-8 (E.M.) and NSF (USA) through Grants No. DMR-1005751 and DMR-1410132 (S.T. and V.D.)

[1] P. Anderson, Science **235**, 1196 (1987).
[2] P. Anderson, P. Lee, M. Randeria, T. Rice, N. Trivedi, and F. Zhang, J. Phys.: Condens. Matter **16**, R755 (2004).
[3] E. Dagotto, Science **309**, 257 (2005).
[4] P. Lee, N. Nagaosa, and X. Wen, Rev. Mod. Phys. **78**, 17 (2006).
[5] C. M. Varma, Comments Solid State Phys. **11**, 221 (1985).
[6] B. J. Powell and R. H. McKenzie, J. Phys.: Condens. Matter **18**, R827 (2006).
[7] B. J. Powell and R. H. McKenzie, Rep. Prog. Phys. **74**, 056501 (2011).
[8] D. C. Johnston, Adv. Phys. **59**, 803 (2010).
[9] K. McElroy, J. Lee, J. A. Slezak, D.-H. Lee, H. Eisaki, S. Uchida, and J. C. Davis, Science **309**, 1048 (2005).
[10] K. Fujita, A. R. Schmidt, E.-A. Kim, M. J. Lawler, D. H. Lee, J. Davis, H. Eisaki, and S.-i. Uchida, J. Phys. Soc. Jap. **81**, 1005 (2012).
[11] B. Keimer, S. A. Kivelson, M. R. Norman, S. Uchida, and J. Zaanen, Nature **518**, 179 (2015).
[12] S. H. Hong, J. M. Bok, W. Zhang, J. He, X. J. Zhou, C. M. Varma, and H.-Y. Choi, Phys. Rev. Lett. **113**, 057001 (2014).
[13] A. Garg, M. Randeria, and N. Trivedi, Nature Phys. **4**, 762 (2008).
[14] N. Fukushima, C.-P. Chou, and T. K. Lee, J. Phys. Chem. Solids **69**, 3046 (2008).
[15] N. Fukushima, C.-P. Chou, and T. K. Lee, Phys. Rev. B **79**, 184510 (2009).
[16] G. G. Guzmán-Verri, A. Shekhter, and C. M. Varma, EPL (Europhysics Letters) **103**, 27003 (2013), URL http://stacks.iop.org/0295-5075/103/i=2/a=27003.
[17] S. Tang, E. Miranda, and V. Dobrosavljevic, Phys. Rev. B **91**, 020501 (2015).
[18] P. Coleman, Phys. Rev. B **29**, 3035 (1984).
[19] G. Kotliar and A. Ruckenstein, Phys. Rev. Lett. **57**, 1362 (1986).
[20] G. Kotliar and J. Liu, Phys. Rev. B **38**, 5142 (1988).
[21] P. A. Lee, N. Nagaosa, T.-K. Ng, and X.-G. Wen, Phys. Rev. B **57**, 6003 (1998).
[22] A. A. Abrikosov and L. P. Gor'kov, Sov. Phys. JETP **12**, 1243 (1961).
[23] See the Supplemental Material at http://link.aps.org/supplemental/XXXXX, which includes references [4, 17, 22, 24, 25], where the AG theory of $T_c$ suppression and normal state resistivity is solved for the disorder potential renormalized by interactions.
[24] K. H. Bennemann and J. B. Ketterson, *Superconductivity: Conventional and unconventional superconductors*, vol. 1 (Springer, 2008).

[25] T. Ando, A. B. Fowler, and F. Stern, Rev. Mod. Phys. **54**, 437 (1982).

# Supplemental Material for "Strong correlations protect $T_c$ against disorder"

Shao Tang,[1] V. Dobrosavljević,[1] and E. Miranda[2]

[1]*Department of Physics and National High Magnetic Field Laboratory,
Florida State University, Tallahassee, Florida 32306, USA*
[2]*Instituto de Física Gleb Wataghin, Campinas State University,
Rua Sérgio Buarque de Holanda, 777, CEP 13083-859, Campinas, Brazil*

### A. The gap equation in AG theory

Within the $U(1)$ slave boson theory for the $t-J$ model of the cuprates, the superconducting transition *in the over-doped regime* is described by the usual BCS equation for a $d$-wave superconductor in which it is the spinons (the auxiliary $f$-fermions) which pair to form the condensate [1]. In other words, the transition is signaled by a non-zero value of the $d$-wave pairing order parameter $\Delta_{ij} = \langle f_{i\uparrow} f_{j\downarrow} - f_{i\downarrow} f_{j\uparrow} \rangle$. We remind the reader that at the transition the other auxiliary fields ($r_i$, $\lambda_i$, $\chi_{ij}$) are well formed. The effects of non-magnetic impurities on $T_c$ can then be treated by using the AG theory [2]. Within that theory, the linearized gap equation is written as [3]

$$\Delta_0 = \frac{2\tilde{J}kT_c}{d}\sum_{i\omega_n}\int \frac{d^2\mathbf{k}}{(2\pi)^2} G^f(\mathbf{k},-i\omega_n) G^f(-\mathbf{k},i\omega_n) \Lambda(\mathbf{k},i\omega_n) \Gamma_d(\mathbf{k}), \quad (1)$$

where $\Delta_0$ is the superconducting gap amplitude, $\omega_n = (2n+1)\pi T_c$, and $\Lambda(\mathbf{k},i\omega_n)$ is the vertex correction function, which satisfies

$$\Lambda(\mathbf{k},i\omega_n) = \Delta_0 \Gamma_d(\mathbf{k}) + n\int \frac{d^2\mathbf{k}'}{(2\pi)^2} |\langle \mathbf{k}| T^f |\mathbf{k}'\rangle|^2 G^f(\mathbf{k}',-i\omega_n) G^f(-\mathbf{k}',i\omega_n) \Lambda(\mathbf{k}',i\omega_n). \quad (2)$$

In the last equation, $\langle \mathbf{k}| T^f |\mathbf{k}'\rangle$ is the single-impurity scattering $T$-matrix for the spinons. The disorder-averaged spinon Green's function in Eqs. (1) and (2) is [2]

$$G^f(\mathbf{k},i\omega_n) = \frac{1}{i\omega_n\left(1+\frac{1}{2\tau_\mathbf{k}|\omega_n|}\right) - \tilde{h}(\mathbf{k}) - \nu_0}, \quad (3)$$

where $\tilde{h}(\mathbf{k}) = -\left(tx + \tilde{J}\chi\right)\Gamma_s(\mathbf{k})$ is the renormalized dispersion, $\nu_0 = \lambda_0 - \mu_0$ is the negative value of the renormalized chemical potential, which controls the doping level, and

$$\frac{1}{\tau_\mathbf{k}} \equiv 2\pi n \int \frac{d^2\mathbf{k}'}{(2\pi)^2} |\langle \mathbf{k}| T^f |\mathbf{k}'\rangle|^2 \delta\left[\tilde{h}(\mathbf{k}) - \tilde{h}(\mathbf{k}')\right], \quad (4)$$

is the quasiparticle scattering rate.

Following the arguments in [2, 3], Eq. (2) can be solved to first order to give

$$\Lambda(\mathbf{k},i\omega_n) = \Delta_0\left[\Gamma_d(\mathbf{k}) + \frac{1}{2\tau_\mathbf{k}^d |\omega_n|\left(1+\frac{1}{2\tau_\mathbf{k}|\omega_n|}\right)}\right], \quad (5)$$

where

$$\frac{1}{\tau_\mathbf{k}^d} \equiv 2\pi n \int \frac{d^2\mathbf{k}'}{(2\pi)^2} |\langle \mathbf{k}| T^f |\mathbf{k}'\rangle|^2 \delta\left(\tilde{h}(\mathbf{k}) - \tilde{h}(\mathbf{k}')\right) \Gamma_d(\mathbf{k}'), \quad (6)$$

is a different scattering rate. We will show later that [see Eq. (33)], when calculated at the approximately circular Fermi surface $\mathbf{k} = k_F \hat{\mathbf{k}}$, the quasiparticle scattering time is essentially isotropic $\tau_\mathbf{k} \approx \tau$, whereas

$$\frac{1}{\tau_\mathbf{k}^d} = \frac{1}{\tau^d}\Gamma_d(\mathbf{k}). \quad (7)$$

We can thus write

$$\Lambda(\mathbf{k}, i\omega_n) = \Delta_0 \Gamma_d(\mathbf{k}) \left[ 1 + \frac{1}{2\tau^d |\omega_n| \left(1 + \frac{1}{2\tau |\omega_n|}\right)} \right]. \tag{8}$$

Eq. (2) can now be solved to all orders by noting that

$$\begin{aligned}\Lambda(\mathbf{k}, i\omega_n) &= \Delta_0 \Gamma_d(\mathbf{k}) \left\{ 1 + \frac{1}{2\tau^d |\omega_n| \left(1 + \frac{1}{2\tau |\omega_n|}\right)} + \left[ \frac{1}{2\tau^d |\omega_n| \left(1 + \frac{1}{2\tau |\omega_n|}\right)} \right]^2 + \ldots \right\} \\ &= \Delta_0 \Gamma_d(\mathbf{k}) \left[ 1 - \frac{1}{2\tau^d |\omega_n| \left(1 + \frac{1}{2\tau |\omega_n|}\right)} \right]^{-1} \\ &= \Delta_0 \Gamma_d(\mathbf{k}) \frac{|\omega_n| + \frac{1}{2\tau}}{|\omega_n| + \frac{1}{2\tau} - \frac{1}{2\tau^d}}. \end{aligned} \tag{9}$$

Plugging this result into Eq. (1), we find that $T_c$ is determined by:

$$1 = \frac{2\widetilde{J}kT_c}{d} \sum_{i\omega_n} \int \frac{d^2\mathbf{k}}{(2\pi)^2} G^f(\mathbf{k}, -i\omega_n) G^f(-\mathbf{k}, i\omega_n) \frac{|\omega_n| + \frac{1}{2\tau}}{|\omega_n| + \frac{1}{2\tau} - \frac{1}{2\tau^d}} \Gamma_d^2(\mathbf{k}). \tag{10}$$

The momentum integral is, as usual, dominated by the region close to the renormalized Fermi surface, which we assume to be approximately circular in the over-doped region. We thus get, using polar coordinates in the $(k_x, k_y)$-plane,

$$1 = \frac{\widetilde{J}kT_c}{d} \sum_{i\omega_n} \int \frac{d\theta}{2\pi} \frac{1}{|\omega_n| + \frac{1}{2\tau} - \frac{1}{2\tau^d}} \Gamma_d^2(k_F \hat{\mathbf{k}}). \tag{11}$$

Using now $\Gamma_d(k_F \hat{\mathbf{k}}) = 2(\cos k_x - \cos k_y) \approx k_x^2 - k_y^2 \approx k_F^2 \cos(2\theta)$ we get

$$1 = \frac{\widetilde{J}kT_c m^* k_F^2}{d} \sum_{n \geq 0} \frac{1}{\omega_n + \frac{1}{2\tau} - \frac{1}{2\tau^d}}. \tag{12}$$

As usual, the integral is formally divergent, but by comparing with the equally divergent expression for the clean transition temperature $T_{c0}$, we can get the ratio of clean ($T_{c0}$) to dirty ($T_c$) transition temperatures [2]

$$\ln \frac{T_{c0}}{T_c} = \psi\left(\frac{1}{2} + \frac{\alpha}{2}\right) - \psi\left(\frac{1}{2}\right), \tag{13}$$

where $\alpha = \frac{1}{2\pi T_c} \left(\frac{1}{\tau} - \frac{1}{\tau^d}\right) \equiv \frac{1}{2\pi T_c \tau_{pb}}$. The leading behavior is

$$T_c = T_{c0} - \frac{\pi}{8\tau_{pb}}. \tag{14}$$

The relevant scattering rates $\tau$ and $\tau_d$ will be calculated in the next Section.

### B. $T$-matrix for the spinons and the pair-breaking scattering rate

As seen in Section A, the suppression of the superconducting transition temperature $T_c$ by disorder requires the determination of the scattering $T$-matrix of the $f$ fermions. We will find it to first order in the disorder. In other words, the fields $(r_i, \lambda_i)$ will be calculated to $\mathcal{O}(\varepsilon_i)$. We define the renormalized site energy for the $f$ electrons as $v_i \equiv \varepsilon_i - \mu_0 + \lambda_i$, whose clean value limit is $v_0 = \lambda_0 - \mu_0$. Clean and disordered $f$-fermion Green's functions are given by, respectively,

$$\mathbf{G}_0^{f-1} = \left[i\omega_n \mathbf{1} + r_0^2 t \mathbf{\Gamma}_s - v_0 \mathbf{1} + \tilde{J}\chi \mathbf{\Gamma}_s\right], \tag{15}$$

$$\mathbf{G}^{f-1} = \left[i\omega_n \mathbf{1} - \mathbf{v} + t\mathbf{r}\mathbf{\Gamma}_s \mathbf{r} + \tilde{J}\chi \mathbf{\Gamma}_s\right], \tag{16}$$

where we have used boldface to denote matrices in the lattice site basis, whose elements are $\mathbf{1}_{ij} = \delta_{ij}$, $\mathbf{r}_{ij} = r_i \delta_{ij}$, $\mathbf{v}_{ij} = v_i \delta_{ij}$, and $(\mathbf{\Gamma}_s)_{ij}$ is equal to 1 if sites $i$ and $j$ are nearest neighbors and zero otherwise. We remind the reader that we are neglecting spatial fluctuations of the $\chi_{ij}$ field. The spinon $T$-matrix is defined through

$$\mathbf{G}^f = \mathbf{G}_0^f + \mathbf{G}_0^f \mathbf{T}^f \mathbf{G}_0^f = \mathbf{G}_0^f \left(1 + \mathbf{T}^f \mathbf{G}_0^f\right), \tag{17}$$

from which we obtain

$$\mathbf{G}^{f-1} = \left(1 + \mathbf{T}^f \mathbf{G}_0^f\right)^{-1} \mathbf{G}_0^{f-1}, \tag{18}$$

and, to first order in the disorder,

$$\mathbf{G}^{f-1} \approx \left(1 - \mathbf{T}^f \mathbf{G}_0^f\right) \mathbf{G}_0^{f-1} = \mathbf{G}_0^{f-1} - \mathbf{T}^f. \tag{19}$$

Thus, again to first order,

$$\begin{aligned} \mathbf{T}^f &= \mathbf{G}_0^{f-1} - \mathbf{G}^{f-1} \\ &= (\mathbf{v} - v_0 \mathbf{1}) - t\mathbf{r}\mathbf{\Gamma}_s \mathbf{r} + r_0^2 t \mathbf{\Gamma}_s \\ &= \delta\mathbf{v} - t r_0 (\delta\mathbf{r}\mathbf{\Gamma}_s + \mathbf{\Gamma}_s \delta\mathbf{r}), \end{aligned} \tag{20}$$

where $\delta\mathbf{r} \equiv (\mathbf{r} - r_0 \mathbf{1})$ and $\delta\mathbf{v} \equiv (\mathbf{v} - v_0 \mathbf{1})$. Defining $\delta v_i \equiv \varepsilon_i + \lambda_i - \lambda_0$ and $\delta r_i = r_i - r_0$, we have, in components,

$$\mathbf{T}^f_{ij} = \delta v_i \delta_{ij} - r_0 (\delta r_i + \delta r_j) t_{ij}. \tag{21}$$

All we need now is to find $\delta r_i$ and $\delta v_i$ to first order in $\varepsilon_i$. This was already obtained in reference [4] (see, in particular, the Supplemental Material). After setting in those equations the gap and its fluctuations to zero $\mathbf{\Delta} = \delta\mathbf{\Delta} = 0$ (normal state) and $\delta\chi = 0$ (as being negligible), we obtain in $\mathbf{k}$-space

$$\Pi^a(\mathbf{k}) \delta v(\mathbf{k}) + r_0 \left[1 + \Pi^b(\mathbf{k})\right] \delta r(\mathbf{k}) = 0, \tag{22}$$

$$\left[\lambda_0 - \frac{\lambda_0}{2d}\Gamma_s(\mathbf{k})\right] \delta r(\mathbf{k}) + r_0 \delta v(\mathbf{k}) = r_0 \varepsilon(\mathbf{k}), \tag{23}$$

where

$$\Pi^a(\mathbf{k}) = \frac{1}{V}\sum_{\mathbf{q}} \frac{f\left[\widetilde{h}(\mathbf{q}+\mathbf{k})\right] - f\left[\widetilde{h}(\mathbf{q})\right]}{\widetilde{h}(\mathbf{q}+\mathbf{k}) - \widetilde{h}(\mathbf{q})}, \tag{24}$$

$$\Pi^b(\mathbf{k}) = \frac{1}{V}\sum_{\mathbf{q}} \frac{f\left[\widetilde{h}(\mathbf{q}+\mathbf{k})\right] - f\left[\widetilde{h}(\mathbf{q})\right]}{\widetilde{h}(\mathbf{q}+\mathbf{k}) - \widetilde{h}(\mathbf{q})} [h(\mathbf{q}+\mathbf{k}) + h(\mathbf{q})], \tag{25}$$

where $d$ is the lattice dimension ($d = 2$, for our purposes), $f(x)$ is the Fermi-Dirac function, $\Gamma_s(\mathbf{k}) = 2(\cos k_x + \cos k_y)$, and $h(\mathbf{k}) = -t\Gamma_s(\mathbf{k})$ is the bare energy dispersion. Solving Eqs. (22)-(23) for $\delta r(\mathbf{k})$ and $\delta v(\mathbf{k})$

$$\delta v(\mathbf{k}) = -x \frac{\Pi(\mathbf{k}) \varepsilon(\mathbf{k})}{\lambda_0 - \frac{\lambda_0}{2d}\Gamma_s(\mathbf{k}) - x\Pi(\mathbf{k})}, \tag{26}$$

$$\delta r(\mathbf{k}) = r_0 \frac{\varepsilon(\mathbf{k})}{\lambda_0 - \frac{\lambda_0}{2d}\Gamma_s(\mathbf{k}) - x\Pi(\mathbf{k})}, \tag{27}$$

where we used $x = r_0^2$ and defined

$$\Pi(\mathbf{k}) \equiv \frac{1+\Pi^b(\mathbf{k})}{\Pi^a(\mathbf{k})}. \tag{28}$$

Therefore

$$\langle \mathbf{k}| T^f |\mathbf{k}'\rangle = \delta v(\mathbf{k}'-\mathbf{k}) + r_0 [h(\mathbf{k}) + h(\mathbf{k}')] \delta r(\mathbf{k}'-\mathbf{k}) \tag{29}$$

$$= x \left[ \frac{h(\mathbf{k}) + h(\mathbf{k}') - \Pi(\mathbf{k}'-\mathbf{k})}{\lambda_0 - \frac{\lambda_0}{2d}\Gamma_s(\mathbf{k}'-\mathbf{k}) - x\Pi(\mathbf{k}'-\mathbf{k})} \right] \varepsilon(\mathbf{k}'-\mathbf{k}). \tag{30}$$

The result in Eq. (30) is general. For the $T_c$ calculation within the AG theory, we only need it for a single impurity [see Eq. (2)]. We therefore set $\varepsilon_i = t\delta_{i,0}$ or $\varepsilon(\mathbf{k}) = t$. We must now plug Eq. (30) into Eqs. (4) and (6). Since the superconducting pairing is mostly affected by the scattering near Fermi surface, we can set $\mathbf{k} = k_F\hat{\mathbf{k}}$ in Eq. (30). For computations, $k_F$ is taken as the magnitude of the Fermi momentum averaged over the approximately circular Fermi surface. We can thus make the following simplifications: $h(\mathbf{k}) + h(\mathbf{k}') = 2E_F$, $\Pi^b(\mathbf{k}) = 2E_F\Pi^a(\mathbf{k})$, $\Pi(\mathbf{k}) = [\Pi^a(\mathbf{k})]^{-1} + 2E_F$, and $\Pi^a(\mathbf{k} - \mathbf{k}') = -\rho^* g_L(y)$, where $y = \frac{|\mathbf{k}-\mathbf{k}'|}{2k_F} = |\sin(\frac{\varphi}{2})|$, $\varphi = \theta - \theta'$ is the angle between $\mathbf{k}$ and $\mathbf{k}'$, $E_F$ is the bare Fermi energy (obtained by solving the mean-field equations with the constraint of an electron filling of $1-x$), the function $g_L(y)$ is defined as [5]

$$g(y) \equiv \begin{cases} 1 & y \leq 1, \\ 1 - \sqrt{1-y^{-2}} & y > 1, \end{cases} \tag{31}$$

and $\rho^* = \frac{m^*}{2\pi}$ is the renormalized (spinon) density of states. Finally, defining

$$g(y) \equiv \frac{t^2}{\{\rho^*\lambda_0 k_F^2 y^2 g_L(y) + x[1-2\rho^* E_F g_L(y)]\}^2}, \tag{32}$$

we obtain

$$\begin{aligned}
\frac{1}{\tau^d(\theta)} &= x^2 \frac{nm^*}{2\pi} \int_0^{2\pi} d\theta' g\left[\left|\sin\left(\frac{\theta-\theta'}{2}\right)\right|\right] \cos(2\theta') \\
&= x^2 \frac{nm^*}{2\pi} \int_0^{2\pi} du\, g\left[\left|\sin\left(\frac{u}{2}\right)\right|\right] \cos[2(\theta-u)] \\
&= \cos 2\theta \left[ x^2 \frac{nm^*}{2\pi} \int_0^{2\pi} du\, g\left[\left|\sin\left(\frac{u}{2}\right)\right|\right] \cos 2u \right] \\
&\equiv \cos 2\theta \frac{1}{\tau^d} = \Gamma_d(\mathbf{k}) \frac{1}{\tau_d}.
\end{aligned} \tag{33}$$

In the last step we dropped the term in $\sin 2\theta$ since this term vanishes after integration in Eq.(1). Analogously,

$$\frac{1}{\tau} = x^2 \frac{nm^*}{2\pi} \int_0^{2\pi} d\theta\, g\left[\left|\sin\left(\frac{\theta}{2}\right)\right|\right], \tag{34}$$

and

$$\frac{1}{\tau_{pb}} = x^2 \frac{nm^*}{2\pi} \int_0^{2\pi} d\theta\, g\left[\left|\sin\left(\frac{\theta}{2}\right)\right|\right] (1-\cos 2\theta). \tag{35}$$

### C. *T*-matrix for the physical electrons and the normal state transport scattering rate

In order to describe transport in the normal state, we must analyze the physical electron scattering $T$-matrix. The calculation is analogous to the one in Section B. The bare and renormalized Green's functions for the physical electrons are given by

$$\mathbf{G}_0^{e-1} = r_0^{-2} \left[ i\omega_n \mathbf{1} + r_0^2 t \boldsymbol{\Gamma}_s - v_0 \mathbf{1} + \widetilde{J}\chi \boldsymbol{\Gamma}_s \right], \tag{36}$$

$$\mathbf{G}^{e-1} = \mathbf{r}^{-1} \left[ i\omega_n \mathbf{1} - \mathbf{v} + t\mathbf{r}\boldsymbol{\Gamma}_s\mathbf{r} + \widetilde{J}\chi \boldsymbol{\Gamma}_s \right] \mathbf{r}^{-1}. \tag{37}$$

Proceeding to first order in the disorder as before yields

$$\begin{aligned} \mathbf{T}^e &= \mathbf{G}_0^{e-1} - \mathbf{G}^{e-1} \\ &= r_0^{-2}\delta\mathbf{v} - 2v_0 r_0^{-3}\delta\mathbf{r} + \widetilde{J}\chi\left(r_0^{-2}\mathbf{\Gamma}_s - \mathbf{r}^{-1}\mathbf{\Gamma}_s\mathbf{r}^{-1}\right) \\ &= x^{-1}\left[\delta\mathbf{v} - 2v_0 r_0^{-1}\delta\mathbf{r} + \widetilde{J}\chi r_0^{-1}\left(\delta\mathbf{r}\mathbf{\Gamma}_s + \mathbf{\Gamma}_s\delta\mathbf{r}\right)\right]. \end{aligned} \tag{38}$$

$$\langle \mathbf{k}|T^e|\mathbf{k}'\rangle = -\left\{\frac{\Pi(\mathbf{k}'-\mathbf{k}) + \frac{2v_0}{x} + \frac{\widetilde{J}\chi}{tx}[h(\mathbf{k}) + h(\mathbf{k}')]}{\lambda_0 - \frac{\lambda_0}{2d}\Gamma_s(\mathbf{k}'-\mathbf{k}) - x\Pi(\mathbf{k}'-\mathbf{k})}\right\}\varepsilon(\mathbf{k}'-\mathbf{k}). \tag{39}$$

This scattering $T$-matrix can now be used to calculate the transport scattering rate that enters the expression for the conductivity in the normal state

$$\frac{1}{\tau_\mathbf{k}^{tr}} \equiv 2\pi x n \int \frac{d^2\mathbf{k}'}{(2\pi)^2}\left|\langle\mathbf{k}|T^e|\mathbf{k}'\rangle\right|^2 \delta\left[\widetilde{h}(\mathbf{k}) - \widetilde{h}(\mathbf{k}')\right](1 - \cos\varphi), \tag{40}$$

where, as before, $\varphi$ is the angle between $\mathbf{k}$ and $\mathbf{k}'$ and the $x$ factor in front of the integral comes from the quasiparticle weight of the physical electron Green's function. We now focus again on wave vectors close to the approximately circular Fermi surface and implement the same approximations used in the previous Section. We first note that the numerators of the fractions is Eqs. (30) and (39), though seemingly different, are actually the same, since

$$\frac{2v_0}{x} + \frac{\widetilde{J}\chi}{tx}[h(\mathbf{k}) + h(\mathbf{k}')] = -2\frac{tx + \widetilde{J}\chi}{tx}E_F + 2\frac{\widetilde{J}\chi}{tx}E_F = -2E_F, \tag{41}$$

where the negative of the renormalized chemical potential $v_0 = -\frac{m}{m^*}E_F = -\frac{tx+\widetilde{J}\chi}{t}E_F$. Like the quasiparticle scattering rate, the transport scattering rate does not depend on the wave vector direction within the assumed approximations and we get

$$\frac{1}{\tau^{tr}} = \frac{xnm^*}{2\pi}\int_0^{2\pi} d\theta\, g\left[\left|\sin\left(\frac{\theta}{2}\right)\right|\right](1 - \cos\theta). \tag{42}$$

---

[1] P. Lee, N. Nagaosa, and X. Wen, Rev. Mod. Phys. **78**, 17 (2006).
[2] A. A. Abrikosov and L. P. Gor'kov, Sov. Phys. JETP **12**, 1243 (1961).
[3] K. H. Bennemann and J. B. Ketterson, *Superconductivity: Conventional and unconventional superconductors*, vol. 1 (Springer, 2008).
[4] S. Tang, E. Miranda, and V. Dobrosavljevic, Phys. Rev. B **91**, 020501 (2015).
[5] T. Ando, A. B. Fowler, and F. Stern, Rev. Mod. Phys. **54**, 437 (1982).